\documentclass[aps,pra,showpacs,twocolumn]{revtex4}
\usepackage{amsfonts}
\usepackage{amssymb}
\usepackage{amsmath}
\usepackage{graphicx}
\usepackage{epsfig}

\input{tcilatex}
\begin{document}

\title{SU(2) symmetry in a Hubbard model with spin-orbit coupling}
\author{X. Z. Zhang, L. Jin, and Z. Song}
\email{songtc@nankai.edu.cn}
\affiliation{School of Physics, Nankai University, Tianjin 300071, China}

\begin{abstract}
We study the underlying symmetry in a spin-orbit coupled tight-binding model
with Hubbard interaction. It is shown that, in the absence of the on-site
interaction, the system possesses the SU(2) symmetry arising from the time
reversal symmetry. The influence of the on-site interaction on the symmetry
depends on the topology of the networks: The SU(2) symmetry is shown to be
the spin rotation symmetry of a simply-connected lattice, so it still holds
in the presence of the Hubbard correlation. In contrary, the on-site
interaction breaks the SU(2) symmetry of a multi-connected lattice.
\end{abstract}

\pacs{71.70.Ej, 71.10.Fd, 03.65.Vf}
\maketitle


\section{Introduction}

The spin-orbit coupling effect as an important mechanism to control spin
dynamics without introducing an external magnetic field \cite{Winkler}, has
received much attention in the context of spintronics and the attempts to
build a spin-transistor since the first proposal by Datta and Das in 1990
\cite{Datta}. Two widely discussed spin-orbit coupling contributions are the
Rashba and the Dresselhaus effects \cite{Rashba,Bychkov,Dresselhaus}. Among
many interesting questions the most important one concerns the underlying
symmetry of this model, which reveals many far-reaching physical
implications that are not obvious at the first glance\textbf{.} A paradigm
example is the SU(2) symmetry discovered by Bernevig \textit{et al.}\ in a
class of spin-orbit coupled models including the model with equal Rashba and
Dresselhauss coupling constants and the Dresselhaus [110] model. This
finding predicted that a spin precession phenomena should be experimentally
observable \cite{Bernevig}. Most of the previous investigations have been
focused on non-interacting systems, while less attention has been paid to
the existence of the electron correlations\textbf{\ }arising from the
Coulomb interaction.

In this article, we study the underlying symmetry in a spin-orbit coupled
tight-binding model, to which only the time reversal symmetry is required.
It is shown that, in the\ absence of the interaction between electrons, the
system possesses the SU(2) symmetry arising from the time reversal symmetry.
Remarkably, we find that the influence of the on-site interaction on the
symmetry depends on the topology of the networks in the following way. This
SU(2) symmetry is shown to be the spin rotation symmetry of a
simply-connected lattice, so it still holds for the case of nonzero on-site
interaction. In contrary, the on-site interaction breaks the SU(2) symmetry
of a multi-connected lattice. Based on the exact solution of a ring system,
our result is demonstrated explicitly.

The article is organized as follows: in Sec. II, we introduce a general
spin-orbit coupled Hubbard Hamiltonian with the time reversal symmetry. In
Sec. III, we first construct the SU(2) operators for an on-site interaction
free system by using the Kramers degeneracy. In Sec. IV we investigate the
influence of the on-site correlation to the SU(2) symmetry. Sec. V is the
conclusion and a short discussion.

\section{Time reversal symmetry}

The Hamiltonian $H$ is written as follows:
\begin{eqnarray}
&&H=H_{T}+H_{U},  \notag \\
&&H_{T}=\sum_{i\neq j}c_{i}^{\dagger }T_{ij}c_{j}+\text{\textrm{H.c.}}%
+\sum_{i}\mu _{i}c_{i}^{\dagger }c_{i},  \label{H} \\
&&H_{U}=\sum_{i}U_{i}n_{i\uparrow }n_{i\downarrow },  \notag
\end{eqnarray}%
where $c_{i}^{\dagger }$ and $c_{i}$ are the creation and annihilation
fermion operators at the $i$th site\ that have two components,

\begin{equation}
c_{i}^{\dagger }=\left( c_{i\uparrow }^{\dagger },c_{i\downarrow }^{\dagger
}\right) \text{, }c_{i}=\dbinom{c_{i\uparrow }}{c_{i\downarrow }}.
\end{equation}%
Here $H_{T}$\ describes the motion of free particles, while $H_{U}$\
represents the on-site interaction of opposite spin electrons. Unlike the
simple Hubbard model\textbf{, }$T_{ij}/\left\vert T_{ij}\right\vert $\textbf{%
\ }is no longer the unit matrix arising from the coupling between momentum
(and/or position) operators and spin operators. In this work, we do not
restrict the model to be in a certain explicit form, we only require it
possessing the time reversal symmetry. Therefore the\textbf{\ }conclusion of
this article is available to the Rashba and Dresselhaus types of spin--orbit
interactions.

The time reversal operator for a spin-$\frac{1}{2}$ particles takes the form%
\begin{equation}
\mathcal{T}=-i\sigma ^{y}K,
\end{equation}%
where $K$ denotes the complex conjugation operator satisfying%
\begin{equation}
K\left( \text{c-number}\right) =\left( \text{c-number}\right) ^{\ast }K,
\end{equation}%
and $\sigma ^{\alpha }$\ $\left( \alpha =x,y,z\right) $\ are the Pauli
matrices. The Hubbard Hamiltonian $H$ possesses the time reversal symmetry
if $\mathcal{K}$\ commutes with all the matrices $T_{ij}$, i.e., $\left[
\mathcal{T}\text{, }T_{ij}\right] =0$ for arbitrary $\left\{ i,j\right\} $.
After a straightforward algebra, we can find that $T_{ij}$\ should have the
form%
\begin{equation}
T_{ij}=t_{ij}\exp \left( i\frac{\theta _{ij}\widehat{n}_{ij}}{2}\cdot
\overrightarrow{\sigma }\right) ,
\end{equation}%
which can be determined by the specific model. Where $t_{ij}$\ is a real
number, $\theta _{ij}$ is an arbitrary angle and $\widehat{n}_{ij}$\ is a
unit vector. Obviously, $T_{ij}/t_{ij}$\ is a unitary matrix and represents
the spin rotation operation.

As well known, the spin operators of the whole system are defined as

\begin{equation}
s^{\alpha }=\overset{N}{\sum_{i=1}}s_{i}^{\alpha },\text{ }s_{i}^{\alpha }=%
\frac{1}{2}c_{i}^{\dagger }\sigma _{\alpha }c_{i},  \label{S_a}
\end{equation}%
which obey the following commutation relations%
\begin{equation}
\left[ s^{\alpha },s^{\beta }\right] =i\varepsilon _{\alpha \beta \gamma
}s^{\gamma },
\end{equation}%
where $\varepsilon _{\alpha \beta \gamma }$\ is the Levi-Civita symbol.

At first, we concentrate on the case of the Hamiltonian of Eq.(\ref{H})
without the spin-orbit interaction. In the absence of the spin-orbit
interaction, $\theta _{ij}=0$, we have
\begin{equation}
\left[ \overrightarrow{s},H_{\theta _{ij}=0}\right] =0,  \label{S com}
\end{equation}%
i.e., the system $H_{\theta _{ij}=0}$\ possesses the SU(2) spin rotation
symmetry. The conservation of $\overrightarrow{s}$\ leads to the\ spin
inversion symmetry along an arbitrary direction, which is in accordance with
the result of the time reversal symmetry.\ According to the general
analysis, the\ spin inversion symmetry is broken when the spin-orbit
interaction is switched on. However, the time reversal symmetry still holds
for nonzero $\theta _{ij}$.\ The objective of this article aims at the
inverse problem, namely: Can the present system acquire a SU(2) symmetry
from the time reversal symmetry? We will find that it is possible under
certain conditions.

\section{On-site interaction free case}

We start with the case of zero $U$, but nonzero $\theta _{ij}$. Due to the
time reversal symmetry of the present system, the Hamiltonian can always be
diagonalized in the form%
\begin{equation}
H_{T}=\sum_{k,\lambda =1,2}\epsilon _{k}f_{k\lambda }^{\dagger }f_{k\lambda
}=\sum_{k}\epsilon _{k}f_{k}^{\dagger }f_{k},  \label{H_T_f}
\end{equation}%
where $\lambda =1,2$ labels the two fold degeneracy, $k$ labels the energy
levels and the corresponding two component fermion operators%
\begin{equation}
f_{k}^{\dagger }=\left( f_{k1}^{\dagger },f_{k2}^{\dagger }\right) ,\text{ }%
f_{k}=\dbinom{f_{k1}}{f_{k2}}.
\end{equation}%
The Kramers degeneracy allows us to construct the operators

\begin{equation}
\digamma ^{\alpha }=\overset{N}{\sum_{k=1}}\digamma _{k}^{\alpha },\text{ }%
\digamma _{k}^{\alpha }=\frac{1}{2}f_{k}^{\dagger }\sigma _{\alpha }f_{k},
\label{F_a}
\end{equation}%
which obey the SU(2) commutation relations%
\begin{equation}
\left[ \digamma ^{\alpha },\digamma ^{\beta }\right] =i\varepsilon _{\alpha
\beta \gamma }\digamma ^{\gamma }.  \label{F_a_F_b}
\end{equation}%
In this sense, Eq. (\ref{H_T_f}) describes a non-interacting Fermi gas of
spin-1/2 particles. The two Kramers degenerate states can service as the two
components of a pseudo spin. Obviously, we have%
\begin{equation}
\left[ \overrightarrow{\digamma },H_{T}\right] =0,  \label{F_H_T}
\end{equation}%
which means that the system $H_{T}$\ possesses a new type of SU(2) symmetry.
Note that the construction of $\left\{ \digamma ^{\alpha }\right\} $\ is not
unique, since any linear transformation of $\left( f_{k1}^{\dagger
},f_{k2}^{\dagger }\right) $\ cannot change the the facts of Eqs. (\ref%
{F_a_F_b}) and (\ref{F_H_T}). In other words, $H_{T}$\ is invariant under
the local rotation of the pseudo spins $\left\{ \digamma _{k}^{\alpha
}\right\} $.

\begin{figure}[tbp]
\includegraphics[ bb=105 190 489 692, width=5.0 cm, clip]{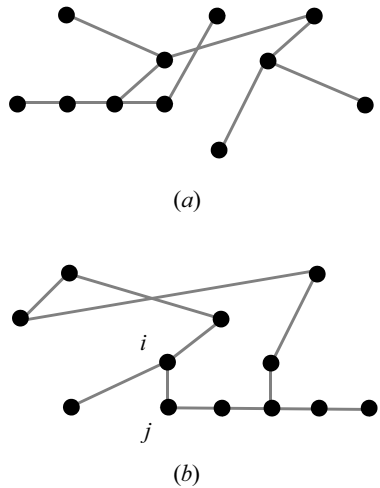}
\caption{(Color online) Schematic illustration of (a) simply-connected
network and (b) multi-connected network. If the connection between $i$ and $%
j $ is broken, the topology is changed from multi- to simply-connected one,
which will affects the SU(2) symmetry of the system, especially in the
presence of the on-site correlation.}
\label{fig1}
\end{figure}

An interesting question is that: Among all the sets of the conserved
quantities $\left\{ \digamma ^{\alpha }\right\} $, which one closes to the
familiar physical quantity and is feasible to measure in experiment. It will
be shown that the geometry and the spin-dependent interaction play an
important role in this issue.\ It is also the main goal of the present
article. To this end, we firstly investigate the Hamiltonian $H_{T}$\ on a
simply-connected network. It can be observed that, taking an arbitrary node
as a start point there exists a unique path to another. A schematic
illustration of the simply-connected network is presented in Fig. \ref{fig1}%
(a). This characteristic feature of such networks allows $H_{T}$ to be
rewritten as

\begin{equation}
\mathcal{H}_{T}=\sum_{i\neq j}t_{ij}d_{i}^{\dagger }d_{j}+\text{\textrm{H.c.}%
}+\sum_{i}\mu _{i}d_{i}^{\dagger }d_{i},  \label{H_T_d}
\end{equation}%
by absorbing the unitary matrices in $T_{ij}$\ into the fermion operators $%
d_{i}^{\dagger }$ and $d_{j}$. For instance, one can take the transformation%
\begin{equation}
c_{i}^{\dagger }T_{ij}c_{j}=t_{ij}c_{i}^{\dagger }e^{i\theta _{ij}\widehat{n}%
_{ij}\cdot \overrightarrow{\sigma }}c_{j}=t_{ij}d_{i}^{\dagger }d_{j},
\end{equation}%
by the definition%
\begin{equation}
d_{i}^{\dagger }=c_{i}^{\dagger },\text{ }d_{j}=e^{i\theta _{ij}\widehat{n}%
_{ij}\cdot \overrightarrow{\sigma }}c_{j}.
\end{equation}%
The equivalent Hamiltonian (\ref{H_T_d}) represents a non-spin-orbit
interaction system. Accordingly, one can construct the corresponding SU(2)
operators
\begin{equation}
\mathcal{S}^{\alpha }=\overset{N}{\sum_{i=1}}\mathcal{S}_{i}^{\alpha },\text{
}\mathcal{S}_{i}^{\alpha }=\frac{1}{2}d_{i}^{\dagger }\sigma _{\alpha }d_{i},
\label{Sigma_a}
\end{equation}%
satisfying the following commutation

\begin{equation}
\left[ \mathcal{S}^{\alpha },\mathcal{S}^{\beta }\right] =i\varepsilon
_{\alpha \beta \gamma }\mathcal{S}^{\gamma }.
\end{equation}%
Similarly with Eq. (\ref{S com}), we have%
\begin{equation}
\left[ \overrightarrow{\mathcal{S}},H_{T}\right] =0.
\end{equation}%
The physics of $\overrightarrow{\mathcal{S}}$ can be understood by the
following relationship between operators $\mathcal{S}_{i}^{\alpha }$\ and $%
s_{i}^{\alpha }$

\begin{equation}
\mathcal{S}_{i}^{\alpha }=u_{i}s_{i}^{\alpha }u_{i}^{^{\dag }},
\label{sigma_S}
\end{equation}%
where $u_{i}$ is a unitary matrix of the form $e^{i\overrightarrow{\gamma }%
_{i}\cdot \overrightarrow{\sigma }}$ appeared in the transformation%
\begin{equation}
d_{i}=u_{i}c_{i}.  \label{duc}
\end{equation}%
This indicates that the operators $\mathcal{S}^{\alpha }$\ act as the real
spin operators under the local transformation $\left\{ u_{i}\right\} $. In
this sense, one can apply the theory of itinerant electron magnetism on the
spin-orbit coupling system on a simply-connected lattice.\ The ferromagnetic
state (all the spins being aligned parallel)\ with respect to the operators $%
\mathcal{S}^{\alpha }$\ in a system with\ nonzero $\theta _{ij}$\ is
equivalent to the spin helix state with respect to the operators $s^{\alpha
} $\ \cite{Bernevig}.\textbf{\ }We will give an extensive discussion about
this issue after taking the on-site interaction into account in the next
section.

On the other hand, the equivalent Hamiltonian (\ref{H_T_d}) can be
diagonalized in the form%
\begin{equation}
H_{T}=\sum_{k,\sigma }\epsilon _{k}d_{k\sigma }^{\dagger }d_{k\sigma },
\label{H_T_dk}
\end{equation}%
where
\begin{eqnarray}
&&d_{k\sigma }=\sum_{j}\mathcal{D}_{j}^{k}d_{j\sigma },  \notag \\
&&\sum_{j}\left( \mathcal{D}_{j}^{k}\right) ^{\ast }\mathcal{D}%
_{j}^{k^{\prime }}=\delta _{kk^{\prime }},  \label{d_k} \\
&&\sum_{k}\left( \mathcal{D}_{j}^{k}\right) ^{\ast }\mathcal{D}_{j^{\prime
}}^{k}=\delta _{jj^{\prime }}.  \notag
\end{eqnarray}%
Comparing Eqs. (\ref{H_T_f}) and (\ref{H_T_dk}), we find that one can have%
\begin{equation}
d_{k}=f_{k}.
\end{equation}%
From Eq. (\ref{d_k}), we obtain the identity

\begin{equation}
\sum_{k}d_{k\sigma }^{\dag }d_{k\sigma ^{\prime }}=\sum_{j}d_{j\sigma
}^{\dag }d_{j\sigma ^{\prime }},
\end{equation}%
which leads to
\begin{equation}
\overrightarrow{\mathcal{S}}=\overrightarrow{\digamma }.
\end{equation}%
Then we can conclude that the SU(2) symmetry obtained from the time reversal
symmetry of a simply-connected system is essentially spin rotational
symmetry. The conservative quantity $\overrightarrow{\digamma }$\ is
connected to an experimental observable $\overrightarrow{\mathcal{S}}$,
which means a persistent spin helix.\textbf{\ }It is hardly observed in
practice since a natural material with simply-connected geometry is rare.
However, artificial lattices,\textbf{\ }such as arrays of quantum\textbf{\ }%
dots in semiconductor heterostructures\textbf{\ }\cite{Lee,Schmidbauer}%
\textbf{\ }confining the conduction electrons, or optical lattices---stable
periodic arrays of potentials created by standing waves of laser light%
\textbf{\ }\cite{Jaksch}, can implement this task.

In contrary, for a multi-connected system illustrated in Fig. \ref{fig1}(b),
the above analysis is invalid since one cannot find a set of unitary
matrices $\left\{ u_{i}\right\} $\ to implement the transformation of Eq. (%
\ref{duc}). Then for an on-site free system with the time reversal symmetry,
it always possesses a SU(2) symmetry, but the physics of the symmetry
depends on the underlying topology of the network.


\section{On-site interaction effect on the symmetry}

In previous we have found that there is a SU(2) symmetry in a spin-orbit
coupling system obeys the time reversal symmetry when the electron-electron
interaction is absent. This indicates the macroscopic emergence of certain
physical features (as ferromagnet, antiferromagnet, spin helix, etc.) in
long time scale, especially in a multi-particle system. Nevertheless, the
spin-dependent interaction between particles may break the symmetry. Now we
turn to investigate the influence of the on-site interaction $H_{U}$ to the
SU(2) symmetry. Actually, applying the transformation $d_{i}=u_{i}c_{i}$ to $%
H_{U}$\ on a simply-connected system, we have%
\begin{equation}
H_{U}=\sum_{i}U_{i}c_{i\uparrow }^{\dagger }c_{i\uparrow }c_{i\downarrow
}^{\dagger }c_{i\downarrow }=\sum_{i}U_{i}d_{i\uparrow }^{\dagger
}d_{i\uparrow }d_{i\downarrow }^{\dagger }d_{i\downarrow },
\end{equation}%
which means it is invariant under the transformation. Consequently for a
simply-connected system, we still have

\begin{equation}
\left[ \overrightarrow{\mathcal{S}},H_{U}\right] =\left[ \overrightarrow{%
\mathcal{S}},H\right] =0.
\end{equation}%
This has many implications on a spin-orbit coupling system. Mathematically,
it can be treated as a normal Hubbard model. Then all the conclusions for
the Hubbard model on a simply-connected lattice are completely available for
the present system. Here we only give a subtle conclusion from the Lieb
theorem \cite{Lieb}. In the following, we present a statement for a Hubbard
model with the spin-orbit interaction on a\ simply-connected network\ by
simply modifying\ the abstract in Ref. \cite{Lieb}. In the attractive
Hubbard Model (and some extended versions of it), the ground state is proved
to have spin angular momentum $\mathcal{S}=0$\ for every (even) electron
filling. In the repulsive case, with a bipartite lattice and a half-filled
band, the ground state has $\mathcal{S}=1/2\left( \left\vert B\right\vert
-\left\vert A\right\vert \right) $, where $\left\vert B\right\vert $\ ($%
\left\vert A\right\vert $) is the number of sites in the $B$\ ($A$)
sublattice. In both cases the ground state is unique. We believe that such
kind of rigorous result obtained from the simple Hubbard model is useful for
understanding the feature of the present model, thereby providing a general
guiding principle for spintronics.

Now we consider the case of a multi-connected system. Since the
transformation $d_{i}=u_{i}c_{i}$ cannot eliminate\ the nonzero $\theta
_{ij} $ term\ completely, we cannot judge the commutation relation $\left[
\overrightarrow{\digamma },H\right] =\left[ \overrightarrow{\digamma },H_{U}%
\right] $ in a general manner.\ However, a single example can provide the
conclusion that the on-site interaction breaks the SU(2) symmetry generated
by the operators $\left\{ \digamma ^{\alpha }\right\} $, although the time
reversal symmetry still holds.

We exemplify the above analysis by taking a simple multi-connected network,
a ring system as an example. This may shed light on the role of the topology
of the network. The Hamiltonian of the ring reads%
\begin{eqnarray}
H^{\text{\textrm{ring}}} &=&H_{T}^{\text{\textrm{ring}}}+H_{U}^{\text{%
\textrm{ring}}},  \label{H_R1} \\
H_{T}^{\text{\textrm{ring}}} &=&-J\sum_{j=1}^{N-1}c_{j}^{\dag
}c_{j+1}-Jc_{1}^{\dag }e^{-i\frac{\pi \sigma _{y}}{2}}c_{N}+\mathrm{H.c.},
\notag \\
H_{U}^{\text{\textrm{ring}}} &=&U\sum_{j=1}^{N}c_{j\uparrow }^{\dagger
}c_{j\uparrow }c_{j\downarrow }^{\dagger }c_{j\downarrow }.  \notag
\end{eqnarray}%
Here the spin-orbit interaction only exerts on the tunneling between sites $%
1 $ and $N$. Despite its simplicity, it reveals the common properties of the
underlying symmetry for more complex systems.\ It acts as a M\"{o}bius
system \cite{Zhao}.\ To obtain the solution, taking the transformation%
\begin{eqnarray}
e^{-i\frac{\left( j-1\right) \pi }{2N}}c_{j\uparrow } &=&a_{j}, \\
e^{-i\frac{\left( N+j-1\right) \pi }{2N}}c_{j\downarrow } &=&a_{N+j},\text{ }%
j\in \left[ 1,N\right] ,  \notag
\end{eqnarray}%
the original Hamiltonian is mapped into
\begin{figure}[tbp]
\includegraphics[ bb=91 183 499 716, width=6.0 cm, clip]{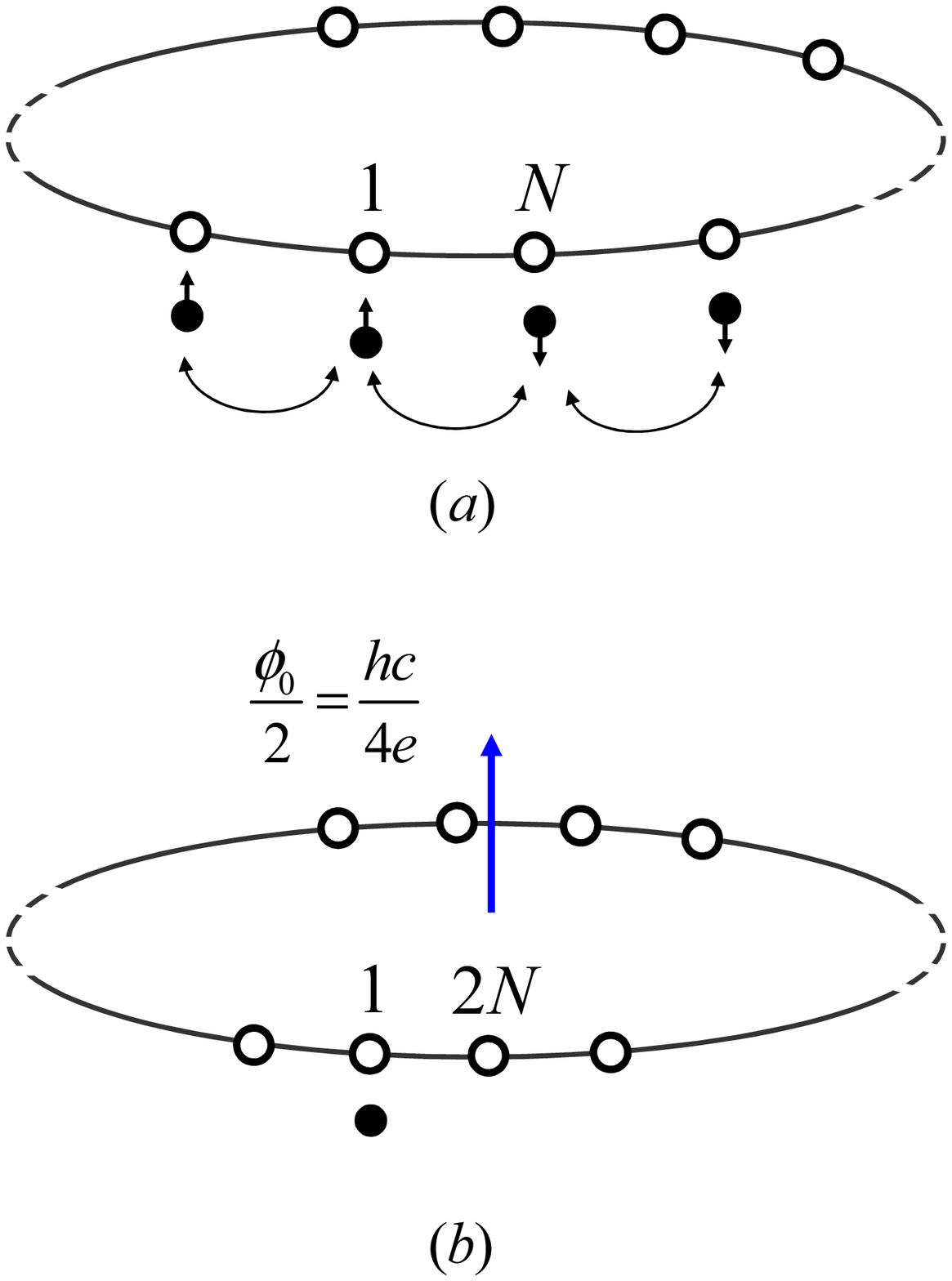}
\caption{(Color online) Schematic illustration of (a) spin-$\frac{1}{2}$
fermionic system of an $N$-site ring with spin-orbit interaction on the bond
between $1$ and $N$, and (b) its equivalent system, which is a spinless
fermionic system of a $2N$-site ring penetrated by a half magnetic flux
quanta.}
\label{fig2}
\end{figure}
%
\begin{equation}
\mathcal{H}_{T}^{\text{\textrm{ring}}}=-J\sum_{j=1}^{2N}\left( e^{i\frac{\pi
}{2N}}a_{j}^{\dag }a_{j+1}+\mathrm{H.c.}\right) ,  \label{H_T_R}
\end{equation}%
where we take the boundary condition $a_{2N+1}^{\dagger }=a_{1}^{\dagger }$.
It represents a $2N$-site ring penetrated by a half magnetic flux quanta,
which is schematically illustrated in Fig. \ref{fig2}. Such a Hamiltonian\
can be diagonalized in the form

\begin{equation}
\mathcal{H}_{T}^{\text{\textrm{ring}}}=-2J\underset{k}{\sum }\cos \left( k+%
\frac{\pi }{2N}\right) a_{k}^{\dagger }a_{k},  \label{H_T_Rk}
\end{equation}%
by using the following Fourier transformation
\begin{equation}
a_{j}=\frac{1}{\sqrt{2N}}\sum_{k}e^{ikj}a_{k},
\end{equation}%
where $k=\pi n/N$, $n\in \lbrack 1,2N]$ denotes the momentum. Note that the
momentum shift by\textbf{\ }$\pi /\left( 2N\right) $ in the dispersion
relation ensures the Kramers degeneracy. Then the corresponding SU(2)
operators (\ref{F_a}) and the on-site interaction $H_{U}^{\text{\textrm{ring}%
}}$\ can be expressed explicitly as%
\begin{eqnarray}
\digamma ^{\alpha } &=&\frac{1}{2}\sum_{k}\left( a_{k}^{\dagger },a_{2\pi
-k-\pi /N}^{\dagger }\right) \sigma _{\alpha }\binom{a_{k}}{a_{2\pi -k-\pi
/N}}, \\
k &=&n\pi /N,\text{ }n\in \left[ 0,N-1\right] ,  \notag
\end{eqnarray}%
and%
\begin{equation}
\mathcal{H}_{U}^{\text{\textrm{ring}}}=U\sum_{j=1}^{N}a_{j}^{\dag
}a_{j}a_{j+N}^{\dag }a_{j+N},
\end{equation}%
respectively. After a lengthy but straightforward algebra, we have

\begin{equation}
\left[ \digamma ^{\alpha },H^{\text{\textrm{ring}}}\right] \neq 0.
\end{equation}%
It is clear that the time reversal symmetry is not the sufficient condition
for a SU(2) symmetry, since the validity of the symmetry depends on the
geometry of the system as well as the on-site correlation. Only the
coexistence of the closed loop in the network and the on-site interaction
between particles can break the SU(2) symmetry.

\section{Conclusion}

In conclusion, we studied the underlying symmetry for a spin-orbit coupled
tight-binding model with the time reversal symmetry. We found that the
characteristics of the symmetry strongly depend on the topology of the
network and the on-site interaction. It is shown that, in the case of zero
on-site interaction, the system possesses the SU(2) symmetry arising from
the Kramers degeneracy. The influence of the on-site interaction on the
symmetry depends on the geometry of the networks: The SU(2) symmetry is
shown to be the spin rotation symmetry of a simply-connected lattice, so it
still holds for the case of nonzero $U$. We also investigate the
multi-connected system based on the exact solution of a simple ring. Our
result showed that the on-site interaction can break the SU(2) symmetry of a
multi-connected lattice.

Regarding the reason why the topology and on-site correlation affects the
symmetry, it may do to its gauge characteristic. It has been pointed that
one can regard the Rashba and the Dresselhaus spin-orbit interaction in
two-dimensional semiconductor heterojunctions as a non-Abelian gauge field,
or the Yang-Mills field \cite{Hatano}. The Yang-Mills field generates a
physical field due to which the wave function acquires a spin-dependent
phase factor. Therefore it is not surprising that the topology of the system
affects the symmetry.

We acknowledge the support of the CNSF (Grant Nos. 10874091 and
2006CB921205).

\end{document}